\documentclass[aps,prl,twocolumn,showpacs]{revtex4}

\usepackage{amsmath,amssymb}
\usepackage{bm}
\usepackage{graphicx}
\usepackage{color}

\newcommand{\braket}[1]{\langle #1 \rangle}

\def\dd{\mathrm{d}}
\def\ee{\mathrm{e}}
\def\ii{\mathrm{i}}

\def\ddt#1{\frac{\partial #1}{\partial t}}

\def\wLP{\omega}
\def\wp{\omega_{\text{p}}}
\def\damprad{\varGamma^{\text{rad}}}
\def\damptot{\varGamma}
\def\nlc{g}

\def\area{S}

\def\ef{\psi}
\def\efd{\psi^{*}}
\def\corrlen{\lambda_\text{c}}

\def\Cin{C^{\text{in}}}
\def\Sout{S^{\text{out}}}

\def\oX{X}
\def\odX{\delta X}

\def\micron{\ensuremath{\mu\mathrm{m}}}

\def\meV{\ensuremath{\mathrm{meV}}}
\def\eV{\ensuremath{\mathrm{eV}}}

\def\vzero{\bm{0}}
\def\vr{\bm{r}}
\def\vk{\bm{k}}
\def\vq{\bm{q}}

\def\oda{\delta\hat{a}}
\def\odad{\delta\hat{a}^{\dagger}}

\def\oa{\hat{a}}
\def\oad{\hat{a}^{\dagger}}

\def\cabpi{\kappa}
\def\dampnr{\varGamma^{NR}}
\def\Nnr{N^{\text{th}}}
\def\oFnr{\hat{F}^{\text{NR}}}
\def\oFnrd{\hat{F}^{\text{NR}\dagger}}

\def\Nact{N_{\text{act}}}
\def\Neff{N_{\text{eff}}}

\begin{document}


\title{Quantum squeezing generation versus photon localization \\ in a disordered microcavity}

\author{Motoaki Bamba}
\altaffiliation{E-mail: motoaki.bamba@univ-paris-diderot.fr}
\author{Simon Pigeon}
\author{Cristiano Ciuti}
\altaffiliation{E-mail: cristiano.ciuti@univ-paris-diderot.fr}
\affiliation{Laboratoire Mat\'eriaux et Ph\'enom\`enes Quantiques,
Universit\'e Paris Diderot-Paris 7 et CNRS, \\ B\^atiment Condorcet, 10 rue
Alice Domont et L\'eonie Duquet, 75205 Paris Cedex 13, France}

\date{\today}

\begin{abstract}
We investigate theoretically the nonlinear dynamics induced by an intense pump field in a disordered planar microcavity. Through a self-consistent theory,  we show how the generation of quantum optical noise squeezing is affected by the breaking of the in-plane translational invariance and the occurrence of photon localization. We find that the  generation of single-mode Kerr squeezing for the ideal planar case can be prevented by disorder as a result of multimode nonlinear coupling, even when the other modes are in the vacuum state. However,  the excess noise is a non-monotonous function of the disorder amplitude. In the strong localization limit,  we show that the system becomes protected with respect to this fundamental coupling mechanism and that the ideal quadrature squeezing generation can be obtained. \end{abstract}


\maketitle

Recently, the intrinsic properties of optically active disordered media have been attracting a considerable interest. In particular, an impressive deal of research has been devoted to the study of the interplay between
lasing and photon localization in random media \cite{Milner2005PRL,Lahini2008PRL,conti08,tureci08,Tulek2010NP},
described in the framework of semiclassical Maxwell-Bloch
equations. Concerning the impact of disorder on purely quantum optical properties, research appears to be in its infancy. Recent works have investigated how an input squeezed light propagate through a linear random medium \cite{smolka09}
or how a slight nonorthogonality of cavity eigenmodes affects the squeezing \cite{lee00}.
The generation of quantum optical squeezing for quantum applications
has been stimulating both in atomic and condensed
matter physics \cite{gardiner04,walls08}.
In the case of an active medium with a third-order optical nonlinearity, it
is  known that the coherent resonant pumping of a photonic mode can produce
a Kerr quantum squeezing of the optical field \cite{fabre97,hilico04}. To understand mechanisms preventing
the ideal squeezing generation is clearly a fundamental issue.
To the best of our knowledge, the  pump-induced generation of quantum squeezing in a nonlinear disordered medium has not been explored.

Here, we wish to explore the role of photon localization
on the quantum squeezing generation of a 2D nonlinear photonic system (like a planar microcavity)
in presence of a disordered potential.
We have consistently determined the nonlinear quantum dynamics of exact
photonic modes in presence of disorder.  Our results shows that the multimode nonlinear coupling
is responsible for an excess noise, which can prevent the quantum squeezing generation. Interestingly we have found that
the excess noise depends non-monotonously on the disorder potential amplitude $\Delta V$. The maximum of noise is obtained for $\Delta V$
comparable to the homogeneous linewidth of the mode. 
However, increasing further the disorder, we find that the ideal quantum squeezing generation can be eventually recovered, as the pumped photon mode becomes "protected" from the multimode coupling in the strong localization limit.


Let us consider a model quantum Hamiltonian for
a 2D photonic system (like a planar microcavity) subject to a third-order nonlinearity and to an in-plane (disordered)
potential for the the photonic motion :

\begin{align} \label{eq:Hamiltonian-V} 
H & = \sum_{\vk} \hbar\omega_{k_z,\vk} \oad_{\vk} \oa_{\vk} +
\sum_{\vk,\vk'} V_{\vk-\vk'} \oad_{\vk} \oa_{\vk'}
\nonumber \\ & \quad
+ \frac{\hbar}{2} \sum_{\vk,\vk',\vq} \nlc_{} \oad_{\vk+\vq} \oad_{\vk'-\vq} \oa_{\vk} \oa_{\vk'},
\end{align}
where $ \omega_{k_z,\vk}  = \frac{c}{\sqrt{\epsilon}} \sqrt{k_z^2+k^2}$ is the frequency of the photonic mode
($\vk$ is the in-plane wavevector, $k_z$ is the quantized vertical wavevector, while $\epsilon$ is the dielectric constant).
$V_{\vk-\mathbf{k'}}$ is the Fourier transform of the in-plane (disordered) spatial potential affecting the in-plane motion of
photons, while $g$ is the nonlinear coupling coefficient for the third-order optical nonlinearity. Moreover, $\oa_{\vk}$ 
is the annihilation operator of a cavity photon with in-plane wavevector $\vk$.

If we diagonalize exactly the Hamiltonian part including the free cavity photon motion and the disordered potential, we can rewrite
the Hamiltonian as:
\begin{equation}
H = \sum_i \hbar\wLP_i \oad_i \oa_i
+ \frac{\hbar}{2} \sum_{ijkl} \nlc_{ijkl} \oad_i \oad_j \oa_k \oa_l,
\end{equation}
where $\wLP_i$ and $\oa_i$ are the eigenfrequency and annihilation boson operator 
of the $i$-th  photon eigenmode in presence of the disorder potential
($i,j,k,l \in \{0,1,2,3,... \}$).
The quantity $\nlc_{ijkl}$ is the nonlinear coupling coefficient depending
on the eigenmode normalized wavefunctions $\ef_i(\vr)$. Namely, it reads:
\begin{equation}
\nlc_{ijkl} = \nlc \int\dd\vr\ \efd_i(\vr) \efd_j(\vr) \ef_k(\vr) \ef_l(\vr).
\end{equation}

A continuous wave coherent optical pump applied to the system is able to drive coherently the intracavity photonic modes and produce a net mean-field $\braket{\oa_i}$. In the Heisenberg picture, the quantum statistical properties of the excited light depend on the fluctuation operator $\oda_i = \oa_i - \braket{\oa_i}$. By linearizing with respect to the fluctuations and by considering the coupling to the extracavity field and to non-radiative reservoirs, we get:
\begin{align} \label{eq:motion-deviation} %
\ii\ddt{}\oda_i & \simeq [\wLP_i-\ii(\damprad_i+\ii\dampnr_i)/2] \oda_i
\nonumber \\ & \quad
+ \sum_{jkl} 2\nlc_{iklj} \braket{\oad_k \oa_l} \oda_j
+ \sum_{jkl} \nlc_{ijkl} \braket{\oa_k \oa_l} \odad_j
\nonumber \\ & \quad
+ \ii\oFnr_i + \ii\ \delta\hat{F}_{i}.
\end{align}
The dynamics of the intracavity fluctuation operator  depends on the  mode eigenfrequency $\omega_i$ and on the radiative loss rate $\damprad_i$ and the non-radiative one $\dampnr_i$.
The terms proportional to $g_{ijkl}$ describe the nonlinear renormalizations and the multimode coupling.
The operator $\hat{F}_i$ is the radiative Langevin force for the $i$-th mode and
$\delta\hat{F}_i \equiv \hat{F}_i - \braket{\hat{F}_i}$,
where $\braket{\hat{F}_i}$ is the excitation amplitude induced
by the coherent optical pumping, where $\braket{\delta\hat{F}_i(t)\delta\hat{F}^{\dagger}_j(t')} 
= \delta_{i,j} \delta(t-t') \damprad_i$.
The non-radiative Langevin force is such that
$\braket{\oFnr_i(t)\oFnrd_j(t')}
= \delta_{ij} \delta(t-t') \dampnr_i(\Nnr_i+1)$, 
where $\Nnr_i$ is the thermal population of $i$-th mode.

The quadrature amplitude of the $i$-th intracavity mode
is defined as 
$\oX^{\rm in}_{i,\theta}(t) = \oa_i(t)\ee^{\ii\wp t-\ii \theta} + \oad_i(t)\ee^{-\ii\wp t+ \ii\theta}$.
The quantum noise is then quantified by the correlation function $\Cin_{i,\theta}(\tau) = \braket{\oX^{\rm in}_{i,\theta}(\tau)\oX^{\rm in}_{i,\theta}(0)}
- \braket{\oX^{\rm in}_{i,\theta}(\tau)}\braket{\oX^{\rm in}_{i,\theta}(0)}
= \braket{\odX^{\rm in}_{i,\theta}(\tau)\odX^{\rm in}_{i,\theta}(0)}$.
The intracavity quadrature noise is given by the equal-time correlation function $\Cin_{i,\theta}(\tau=0)$.
The extracavity quadrature is defined as $\oX^{\rm out}_{i,\theta}(t) = \hat{\alpha}^{\rm OUT}_i(t)\ee^{\ii\wp t-\ii \theta} + \hat{\alpha}^{{\rm OUT}\dagger}_i(t)\ee^{-\ii\wp t+ \ii\theta}$,
where $\hat{\alpha}^{\rm OUT}_i(t)$ is the output operator with the same transverse profile of the $i$-th intracavity mode (see Supplementary Materials).
The extracavity quantum noise is quantified by the Fourier transform of correlation function $\Sout_{i,\theta}(\omega) = \int_{\infty}^{\infty}\dd\tau\ \ee^{\ii\omega\tau}[\braket{\oX^{\rm out}_{i,\theta}(\tau)\oX^{\rm out}_{i,\theta}(0)}
- \braket{\oX^{\rm out}_{i,\theta}(\tau)}\braket{\oX^{\rm out}_{i,\theta}(0)}]$.
For a normal vacuum or for a coherent state, $\Cin_{i,\theta}(\tau=0) = \Sout_{i,\theta}(\omega=0) = 1$ (standard quantum limit).
The intracavity and extracavity squeezing are obtained when $\Cin_{i,\theta}(\tau=0) < 1$ and $\Sout_{i,\theta}(\omega=0) < 1$, respectively.

In the case of a perfectly planar cavity (no disorder), the
eigenfunctions are planewaves $\ef_j(\vr) = \ee^{\ii\vk_j\cdot \vr} /
\sqrt{\area}$  and  the nonlinear coefficients are $\nlc_{ijkl} = \nlc
\delta_{\vk_i+\vk_j,\vk_k+\vk_l}\ $.
If we consider the optically pumping of  the cavity by a resonant laser with in-plane wavevector
$\vq = \vzero$,
only the ground intracavity mode ($\vk_0 = \vzero$) is coherently excited,
and then Eq.~\eqref{eq:motion-deviation} is reduced to
\begin{align} \label{eq:motion-deviation-planar} 
\ii\ddt{}\oda_0 & \simeq 
\left[ \wLP_0-\ii(\dampnr_0+\damprad)/2 
     + 2\nlc \sum_{i} \braket{\oad_i \oa_i} \right] \oda_0
\nonumber \\ & \quad
+ \nlc \braket{\oa_0 \oa_0} \odad_0
+ \ii\oFnr_0 + \ii\ \delta\hat{F}_{0}.
\end{align}
Therefore, since Eq.~\eqref{eq:motion-deviation-planar} 
has no interaction with the fluctuations of the other modes,
the dynamics reduce to that of the well-known single-mode Kerr model. In the case of negligible non-radiative damping and no thermal population ($\Nnr_i =0$), the minimum quantum noise is 
$\Cin_0 \equiv \min[\Cin_{0,\theta}(0)]= 0.5$ and $\Sout_0 \equiv \min[\Sout_{0,\theta}(0)] = 0$, obtained for pumping intensities close to the bistability turning point \cite{yurke84,collett84,collett85,karr04,courty96pra}.
It is known that such extracavity perfect squeezing is due to an interference effect between the incident pump field and the reemitted light.

In presence of disorder, the situation is qualitatively different. We have solved Eq.~\eqref{eq:motion-deviation} by considering a random potential with amplitude $\Delta V$ and correlation length $\corrlen$.
We have calculated the squeezing of the ground intracavity mode (denoted by $i=0$)
assuming that only the ground mode is coherently excited
($\langle \hat{F}_i \rangle = \delta_{i,0}F^{pump}_0(t)$).
This means that only the ground mode has a coherent amplitude
$\braket{\oa_0}$
and that the expectation values in Eq.~\eqref{eq:motion-deviation} are given by
$\braket{\oad_i\oa_j} = \delta_{i,0}\delta_{j,0}|\braket{\oa_0}|^2 + \delta_{i,j}\Nnr_i$ and
$\braket{\oa_i\oa_j} = \delta_{i,0}\delta_{j,0}\braket{\oa_0}^2$.
Following Eq.~\eqref{eq:motion-deviation}, the ground mode ($i=0$) is coupled to the higher modes via the terms
$2\nlc_{000j} \braket{\oad_0 \oa_0} \oda_j$ and
$\nlc_{0j00} \braket{\oa_0 \oa_0} \odad_j$. Such intermode coupling terms produce
an excess noise for the ground photonic mode.

\begin{figure}[tbp] 
\includegraphics[width=\linewidth]{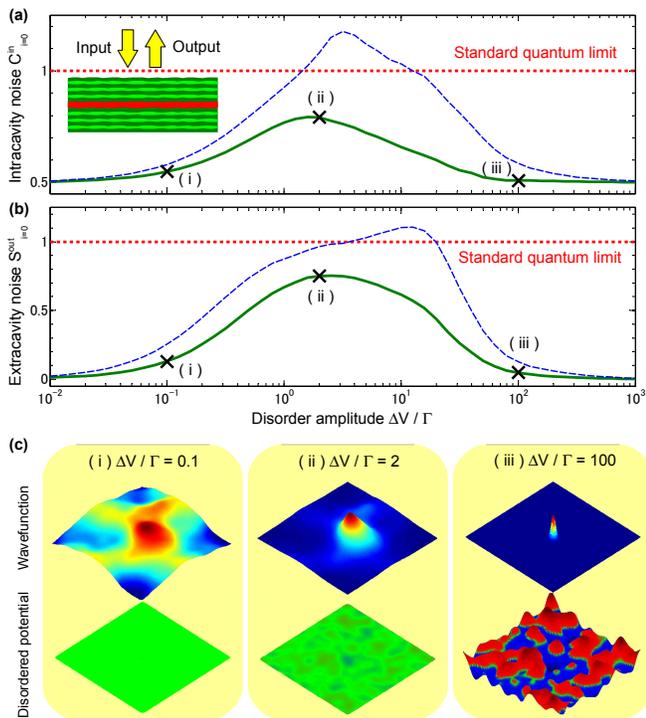}
\caption{(a) Minimum value $\Cin_{i=0}$ of the quadrature quantum optical noise of the ground intracavity mode versus the disorder potential amplitude $\Delta V$ (normalized to the mode homogeneous broadening $\damptot$).
The dotted line ($\Cin_{i=0} = 1$) is the standard quantum limit.
The solid line is calculated under the condition that
only the ground mode is driven coherently,
while all the other modes are in the vacuum state.
The dashed line is obtained by considering a thermal population $N^{\text{th}}_{i\neq 0} = 1$,
while only the ground mode is coherently pumped.
(b) Minimum value $\Sout_{i=0}$ of the quadrature noise of the extracavity field which has the same transverse profile as the ground intracavity mode.
(c) Surface plots of the disordered potential (bottom) and of the ground mode photonic wavefunctions (top) for the different amplitudes (i), (ii), (iii) of the disorder potential, also indicated in (a) and (b).  
Parameters: $100 \micron \times 100 \micron$ system with periodic boundary conditions; disorder potential with correlation length $\corrlen = 5\;\micron$; $\hbar\omega_{k_z,0} = 1.5\;\eV$; $\epsilon = 13$; $g=0.01\;\meV\micron^2$; $\hbar\damptot = \hbar\damprad_0 = \hbar\dampnr_{i\neq0} = 0.1\;\meV$; $\hbar\damprad_{i\neq0}=\hbar\dampnr_{0}=0$; $\wp = \omega_{i=0}+\damptot$.  Averaging over $100$ disorder configurations.
\label{fig:1} }
\end{figure}

Representative results are depicted in Fig.~\ref{fig:1}.
We have considered parameters for a GaAs microcavity (details in the caption). For the case $\Nnr_i = 0$ (no thermal population), the only relevant damping parameter for the intracavity dynamics is the total
 loss rate $\damprad_i + \dampnr_i  = \damptot$. Of course, for the extracavity noise statistics, the ratio $\dampnr_0/ \damprad_0 $ is crucial. In the calculation of Fig.~\ref{fig:1}, we have considered   $\damprad_0  \gg \dampnr_0 $ (ideal condition). In the case $\Nnr_i > 0$, the non-radiative loss rate plays an important role.
For the modes $i \neq 0$, we have considered for simplicity $\dampnr_{i\neq0} = \damptot$. 
We have considered a finite-size 2D system ($100\;\micron\times100\;\micron$)
with periodic boundary conditions and the correlation length of disorder is $\corrlen = 5\;\micron$. The pump frequency is $\wp=\omega_{i=0} + \damptot$.
In Figs.~\ref{fig:1}(a) and \ref{fig:1}(b), we present results for the minimum intracavity quadrature noise $\Cin_{i=0}$ and extracavity one $\Sout_{i=0}$, respectively, as a function of the disorder potential amplitude $\Delta V$, normalized
to the homogeneous broadening $\damptot$.
The solid line refers to the case where $\Nnr_{i\neq 0} = 0$, i.e., the non-pumped modes are in the vacuum state
with no excess thermal noise.
For negligible disorder ($\Delta V/ \damptot \ll 1$), the minimum intracavity noise takes the value $\Cin_{i=0}=0.5$, and the extracavity noise is $\Sout_{i=0}=0$. 
For  $\Delta V/ \damptot =  0.1$ (point (i)) the intracavity and extracavity noises are only slightly increased to $0.55$ and $0.13$, respectively.
Increasing $\Delta V/\damptot$, the minimum noises significantly increase and reach a maximum values of $\Cin_{i=0}=0.77$ and $\Sout_{i=0}=0.75$ for $\Delta V/\damptot = 2$ (point (ii)). Increasing $\Delta V/\damptot$ further, the noises decrease and recover the values $\Cin_{i=0}=0.5$ and $\Sout_{i=0}=0$ (for example, $\Delta V/\damptot = 100$, point (iii)).
 In Fig.~\ref{fig:1}(c), we show a surface representation  
of the photonic ground mode squared wavefunction $|\psi_0(\vr)|^2$ (top)
and of the disordered potential (bottom), showing the increasing degree of localization of the photonic ground wavefunction with 
increasing value of $\Delta V$.
Note that the dashed line in Figs.~\ref{fig:1}(a) and \ref{fig:1}(b) instead is for the case with $\Nnr_{i\neq 0} = 1$, meaning that the non-pumped modes have excess noise due to a thermal population. In such a case, there is a similar behavior to the case $\Nnr_{i\neq 0} = 0$ (solid line), but, importantly, the minimum noise can be larger than $1$ (larger than the standard quantum limit).

\begin{figure}[tbp] 
\includegraphics[width=\linewidth]{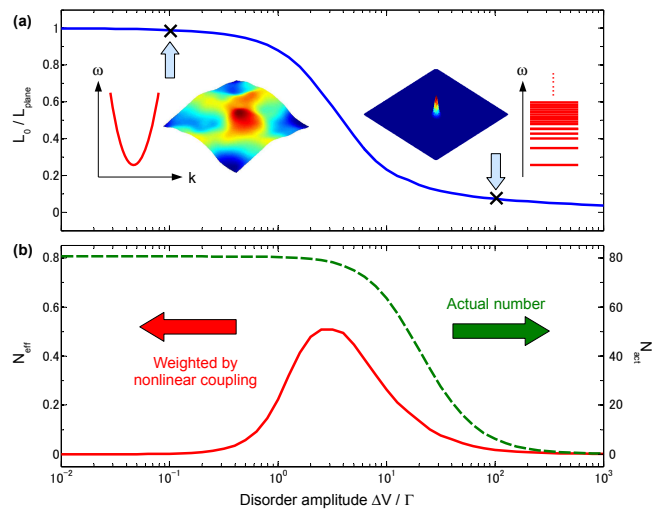}
\caption{(a) Localization length $L_0$ of the ground mode normalized to the system size $L_{\text{plane}}$ versus the normalized disorder amplitude $\Delta V/\damptot$.
(b) Solid line: effective number $\Neff = \pi\damptot W_{\text{DOS}}(\omega_0) - 1$ of modes weighted by the nonlinearity to the ground mode.
Dashed line: actual number $\Nact = \pi\damptot\rho_{\text{DOS}}(\omega_0)-1$ of modes, by taking into account the normal density of states.}
\label{fig:2}
\end{figure}

In order to understand the origin of this interesting behavior of the nonlinear quantum squeezing generation in presence of disorder, we have studied in detail the localization properties, the density of states of the disordered eigenstates and the nonlinear coupling between them. 
We have found that the generation of quantum squeezing can be forbidden if (a) there is a significant number of modes close in energy to the ground mode ($\omega_{i \ge 1} - \omega_0 \sim\damptot$) and (b) they are significantly coupled to the ground mode ($g_{000i}$  has to be comparable to $g_{0000}$).
In order to quantify the localization,  we have considered the localization length
$L_0 = \sqrt{\braket{|\vr-\braket{\vr}|^2}}$ of the photonic ground mode, where the average $\braket{\cdots}$ is taken on the ground mode state.
In Fig.~\ref{fig:2}(a), we plot $L_0$ divided by $L_{\text{plane}}$
(the size of a planewave mode) as a function of disorder amplitude $\Delta V$. For small values of $\Delta V$, $L_0/L_{\text{plane}} \simeq 1$ and,
since the translational invariance is broken only very weakly,  the nonlinear coupling coefficients are such that $\nlc_{000i}/\nlc_{0000} \simeq \delta_{i,0}$. 
However, increasing $\Delta V$, the wavefunction becomes more and more localized and a transition from $L_0/L_{\text{plane}} = 1$ to $L_0/L_{\text{plane}} \ll 1$ occurs.  In the strong localization limit, we have found that an energy gap between the ground photonic mode and the higher modes is opened (see the sketches of energy structures in Fig.~\ref{fig:2}(a)). This energy gap effectively decouples the ground mode ($i=0$) from the other modes ($i \ge 1$) and protects the quantum squeezing generation from the multimode coupling. Hence, the ideal values for single-mode Kerr squeezing are recovered both in the case of negligible and very strong localization, albeit for very different reasons. In the case of no disorder, the selection rule due to the translational invariance is responsible for the mode decoupling; in the strong localization limit, the decoupling is due to a frequency gap $\omega_{i \ge 1} - \omega_0 \gg \damptot$. In between, the multimode coupling produces an excess noise, which is maximum for a critical value of $\Delta V/\damptot$ (around $2$ for the considered type of disorder).

In order to get further insight into the nonlinear multimode coupling, we have considered the density of photonic states $\rho_{\text{DOS}}(\omega)  
= \sum_i \mathcal{L}_i(\omega)$, where $\mathcal{L}_i(\omega)$ is a normalized  Lorentzian function ($\int\dd\omega\ \mathcal{L}_i(\omega) = 1$) 
with center $\omega_i$ and width $\damprad_i+\dampnr_i$. Having the density of states $\rho_{\text{DOS}}(\omega)$, it is possible to compute the number 
$\Nact = \pi\damptot\rho_{\text{DOS}}(\omega_0)-1$ of modes whose eigenfrequencies  differ from $\omega_0$ less than $\damptot$. Such a number $\Nact$ is plotted in Fig.~\ref{fig:2}(b) (dashed line).
It is apparent that $N_{\text{act}}$ is not responsible alone for the non-monotonous
dependence of the quantum noise versus disorder amplitude in Fig.~\ref{fig:1}(a). 
Indeed, as apparent from Eq.~\eqref{eq:motion-deviation}, the Kerr quantum squeezing is generated by the term  $g_{0000}  \langle \oa_0\oa_0 \rangle \delta \oa_0^{\dagger}$,
while the excess noise due to the coupling to the $i$-th mode ($i \neq 0$) comes from the term $g_{0i00}  \langle \oa_0\oa_0\rangle \delta \oa_i^{\dagger}$.
Hence, the ratio between the amplitudes $g_{000i}$ and $g_{0000} $ plays an important role. 
It is therefore instructive to study the behavior of  the quantity
\begin{equation}
W_{\text{DOS}}(\omega)
= \sum_i |\nlc_{000i}/\nlc_{0000}|^2 \mathcal{L}_i(\omega).
\end{equation}
$W_{\text{DOS}}(\omega)$ is a modified density of states where the contribution of  each state is weighted by its nonlinear coupling  to the ground mode. 
Such simple weighted density of states indeed qualitatively reproduces the non-monotonous dependence of the excess noise as a function of the disorder amplitude in Fig.~\ref{fig:1}(a). 
This is confirmed by looking at the effective number of modes obtained with the weighted density of states, namely $\Neff = \pi \damptot W_{\text{DOS}}(\omega_0)-1$, which is plotted by a solid line in Fig.~\ref{fig:2}(b).


From the above discussion, it is now clear that,
in planar cavities with disorder amplitude $\Delta V$ comparable to the mode homogeneous broadening $\damptot$, the generation of quantum squeezing can be severely limited by nonlinear coupling between disordered states. This critical condition $\Delta V \simeq \damptot$ is actually achieved in semiconductor GaAs microcavities, which have been experimentally investigated in recent works \cite{karr04,deveaud07,bajoni08}.
However, our results also show that the system where photonic energy levels are separated much more than the homogeneous broadening $\damptot$  the ideal single-mode Kerr squeezing can be recovered. Therefore, in this context, it appears natural to consider artificially localized photonic systems such as micropillars \cite{bajoni08} to achieve a protected quantum squeezing generation. In GaAs-based microcavities, discrete photon levels are obtained already in a micropillar
with a $4\;\micron{}$ diameter and a level separation larger than  the broadening can be achieved \cite{note}.
Of course, the complete suppression of extracavity noise can be detected only by using a local oscillator \cite{fabre97} with the same spatial shape as the ground photonic wavefunction.


In conclusion, we have presented a theory for the role of photon localization on the nonlinear generation of quantum squeezing in a disordered planar microcavity. We have shown that the ideal single-mode quantum noise reduction is limited by nonlinear multimode coupling between disordered eigenmodes, even when the modes are in the vacuum state. Protection from multimode coupling is obtained in the strong localization limit \cite{note2}. Our study shows the interesting link between nonlinear quantum noise properties and photon localization and may stimulate a wide range of experimental and theoretical investigations in a variety of systems.
For example, the link between Anderson localization and quantum noise in 3D photonic systems \cite{conti08}, or the quantum noise statistics of superfluid light in a nonlinear disordered medium \cite{carusotto04,amo09naturephys} are intriguing problems to explore in the future.

\begin{acknowledgments}
We wish to thank A. Amo, A. Bramati, I. Carusotto, C. Fabre, E. Giacobino, J. Ph. Karr for  discussions
and/or for a critical reading of the manuscript.
\end{acknowledgments}

\section{Supplementary material}
\subsection{1. Calculation of in-plane eigenmodes}
As disordered potential $V_{\vk}$ for the in-plane motion of photons, we have considered a Gaussian function \cite{savona97}  with amplitude $\Delta V$ and correlation length $\corrlen$. Namely, we have taken 
$V_{\vk} 
= \Delta V \ee^{-{\corrlen}^2k^2/4 + \ii\theta_{\mathrm{k}}}$,
where the phase $\theta_{\mathrm{k}}$ is given by a random function
satisfying $\theta_{\vk} = - \theta_{-\vk}$
to make 
$V(\vr) = \sum_{\vk} V_{\vk} \ee^{\ii\vk\cdot\vr}$
real.
The 2D system with size $L_x \times L_y$ is divided into a $N_x \times N_y$ grid ($N_x=N_y=128$ in our numerical calculations)
and the in-plane wavevector $\vk = (k_x, k_y)$
in the first Brillouin zone is used: $k_{x,y} = 2\pi n/L_{x,y}$
and $n = -N_{x,y}/2, -N_{x,y}/2+1, \ldots, N_{x,y}/2-1$.
The eigenfrequency $\omega_i$ and function $\psi_i(\vr)$
in the 2D system is calculated by solving
a eigenvalue problem
$[\hbar\omega_{k_z}(-\ii\nabla) + V(\vr)] \psi_i(\vr)
= \hbar\omega_i \psi_i(\vr)$, where
$\omega_{k_z}(-\ii\nabla) = \frac{c}{\sqrt{\epsilon}}\sqrt{{k_z}^2 - \nabla^2}$.
We have used periodic boundary conditions for the photonic wavefunction. In the numerical diagonalization, the matrix is relatively sparse.

\subsection{2. Treatment of extracavity fields}
For the treatment of extracavity fields in Eq.~(4) on the main paper,
we have followed a generalized input-output approach for the present system. The coupling between the intracavity and the extracavity
fields is described by the following Hamiltonian:
\begin{equation} \label{eq:Hextra} 
H_{\text{extra}}
= - \frac{\ii\hbar}{\sqrt{2\pi}} \int_{-\infty}^{\infty}\dd\omega
  \sum_{i,\vq}
  \left[ \cabpi_{i,\vq} \oad_i \hat{\alpha}_{\vq}(\omega)
       - \cabpi_{i,\vq}^* \hat{\alpha}_{\vq}^{\dagger}(\omega) \hat{a}_i \right].
\end{equation}
$\hat{\alpha}_{\vq}(\omega)$ is 
the annihilation operator of the extracavity field
with frequency $\omega$ and in-plane wavevector $\vq$
and the coupling coefficient $\cabpi_{i,\vq}$ is represented
by the Fourier transform $\tilde{\psi}_{i,\vk}$ of eigenfunction $\psi_i(\vr)$ as
$\cabpi_{i,\vq} = \sqrt{\damprad} \tilde{\psi}_{i\vk}$.
Here, it is convenient to consider a set of extracavity fields with the same
transverse profile as the intracavity eigenmodes, namely:
\begin{equation}
\hat{\alpha}_i(\omega) \equiv \sum_{\vq} \hat{\alpha}_{\vq}(\omega) \tilde{\psi}_{i,\vq}.
\end{equation}
In terms of this basis, Eq.~\eqref{eq:Hextra} is rewritten as
\begin{equation}
H_{\text{extra}}
= -\ii\hbar \sqrt{\frac{\damprad}{2\pi}} \int_{-\infty}^{\infty}\dd\omega
  \sum_{i}
  \left[ \oad_i \hat{\alpha}_{i}(\omega)
       - \hat{\alpha}_{i}^{\dagger}(\omega) \hat{a}_i \right].
\end{equation}
Therefore, the extracavity mode with the same transverse spatial profile as the $i$-th intracavity mode is not affected by the other orthogonal intracavity modes, and then Eq.~(4) on the main paper and the input-output relation
\begin{equation}
 \hat{\alpha}^{\rm OUT}_{i}(\omega)
= - \hat{\alpha}^{\rm IN}_{i}(\omega) + \sqrt{\damprad} \hat{a}_{i}(\omega)
\end{equation}
are derived by the standard approach \cite{gardiner04,walls08}.
$\hat{\alpha}^{\rm IN/OUT}_{i}(\omega)$ is the so-called input/output operator (operator $\hat{\alpha}_{i}(\omega,t_0)$ where $t_0 \to -\infty/\infty$) is the initial/final time of the dynamics, and the Langevin force operator in Eq.~(4) on the main paper is written as
\begin{equation}
\hat{F}_i(t) = \braket{\hat{F}_i(t)} + \delta\hat{F}(t)
= \frac{1}{\sqrt{2 \pi}} \int_{-\infty}^{\infty} \dd\omega\
\ee^{-\ii \omega t} \hat{\alpha}^{\rm IN}_{i}(\omega).
\end{equation}
In practice, the dynamics of extracavity field $\hat{\alpha}_i(\omega)$ can be measured by using a local oscillator spot with the appropriate shape, if the wavefunction $\tilde{\psi}_{i,\vq}$ is mostly concentrated within the light cone. Actually, in our calculation, there is negligible amplitude of the fundamental mode $\tilde{\psi}_{0,\vq}$ outside the light cone.

\subsection{3. Details about correlation functions}
The quadrature amplitude of the $i$-th intracavity mode
is defined as 
$\oX^{\rm in}_{i,\theta}(t) = \oa_i(t)\ee^{\ii\wp t-\ii \theta} + \oad_i(t)\ee^{-\ii\wp t+ \ii\theta}$.
The quantum noise is then quantified by the correlation function $\Cin_{i,\theta}(\tau) = \braket{\oX^{\rm in}_{i,\theta}(\tau)\oX^{\rm in}_{i,\theta}(0)}
- \braket{\oX^{\rm in}_{i,\theta}(\tau)}\braket{\oX^{\rm in}_{i,\theta}(0)}
= \braket{\odX^{\rm in}_{i,\theta}(\tau)\odX^{\rm in}_{i,\theta}(0)}$,
where $\odX^{\rm in}_{i,\theta} = \oX^{\rm in}_{i,\theta} - \braket{\oX^{\rm in}_{i,\theta}}$.
The intracavity quadrature noise is given by the equal-time correlation function $\Cin_{i,\theta}(\tau=0)$.
The extracavity quadrature is defined as $\oX^{\rm out}_{i,\theta}(t) = \hat{\alpha}^{\rm OUT}_i(t)\ee^{\ii\wp t-\ii \theta} + \hat{\alpha}^{{\rm OUT}\dagger}_i(t)\ee^{-\ii\wp t+ \ii\theta}$. Its quantum noise is quantified by the Fourier transform of correlation function $\Sout_{i,\theta}(\omega) = \int_{\infty}^{\infty}\dd\tau\ \ee^{\ii\omega\tau}[\braket{\oX^{\rm out}_{i,\theta}(\tau)\oX^{\rm out}_{i,\theta}(0)}
- \braket{\oX^{\rm out}_{i,\theta}(\tau)}\braket{\oX^{\rm out}_{i,\theta}(0)}]$.
As already well explained in the literature \cite{yurke84,collett84,collett85}, the extracavity field quantum noise properties are affected by the interference between the incident (input) field and the reemitted light by the cavity. If the quadrature noise of the fundamental intracavity mode is $\Cin_{i=0}=0.5$, it can be shown that a perfect squeezing ($\Sout_{i=0}=0$) can be obtained outside in the case $\dampnr_0 \ll \damprad_0$.

\subsection{4. Linearization Bogoliubov method}
Concerning the stability of the mean-field solutions  $\braket{\hat{a}_i}$, we note that 
the brackets in Eq.~(4) on the main paper are calculated as
$\braket{\oad_i\oa_j} = \delta_{i,0}\delta_{j,0}|\braket{\oa_0}|^2 + \delta_{i,j}\Nnr_i$ and
$\braket{\oa_i\oa_j} = \delta_{i,0}\delta_{j,0}\braket{\oa_0}^2$. Moreover, 
we treat $|\braket{\oa_0}|^2$ as a variable
proportional to the pump intensity
and check the system stability by a linearization Bogoliubov method.
When we consider $N$ eigenmodes of in-plane photonic motion,
the size of the Bogoliubov matrix $\bm{\mathsf{L}}$ is $2N \times 2N$.
When $\bm{\mathsf{L}}$ is divided into $N\times N$ blocks with each size
$2\times2$, the $(i,j)$ block is written as
\begin{widetext}
\begin{equation}
\begin{pmatrix}
(\delta\omega-\ii\frac{\damprad_i+\dampnr_i}{2})\delta_{ij}+2g_{i00j}|\braket{\oa_0}|^2 &
g_{ij00}\braket{\oa_0}^2 \\
-g_{ij00}^*\braket{\oa_0}^{*2} &
(-\delta\omega-\ii\frac{\damprad_i+\dampnr_i}{2})\delta_{ij}-2g_{i00j}^*|\braket{\oa_0}|^2
\end{pmatrix},
\end{equation}
\end{widetext}
where $\delta\omega \equiv \wLP_i+2g_{i00i}\Nnr_{i}-\wp$.
The system is stable when the imaginary part of its eigenvalues
are all negative.
In order to find the minimum of intracavity and extracavity quantum noise, 
we calculated the values $\Cin_{i=0}=\min[\Cin_{i=0,\theta}(0)]$
and $\Sout_{i=0}=\min[\Sout_{i=0,\theta}(0)]$
obtained by tuning the quadrature phase $\theta$ and 
by tuning the pump intensity in the stability region.
For the system with only one photon mode, the minimum values $\Cin_{i=0}=0.5$ and $\Sout_{i=0}=0$ are obtained at two pump intensities at an instability boundary.
However, for the system with many modes and no translational invariance, in general, the system has many instability regions 
and the minimum is obtained not just at the boundaries.
We considered the minimum values below the first instability.


\begin{thebibliography}{10}
\bibitem{note}
The microcavity systems in Refs.~\onlinecite{karr04} and \onlinecite{bajoni08}  have been investigated to determine properties of polariton excitations (photon mode strongly coupled to an excitonic transition). The same kind
of system can be used to obtain photon modes with an effective Kerr nonlinearity by properly detuning the frequency of the cavity mode from the exciton level.
\bibitem{note2}
Note that the localization lengths considered in our manuscript are not small enough
to enter the regime of photon (polariton) quantum blockade \cite{verger06} leading
to single-photon states. The squeezing generation in the present manuscript is for
the continuous variable regime (many photons in the pump mode).
\end{thebibliography}

\end{document}